\documentclass[12pt,english]{article}
\usepackage[T1]{fontenc}
\usepackage[latin9]{inputenc}
\usepackage{geometry}
\usepackage{color}
\usepackage{array}
\usepackage{multirow}
\usepackage{amsmath}
\usepackage{amssymb}
\usepackage{graphicx}
\usepackage{esint}

\makeatletter

\providecommand{\tabularnewline}{\\}

\newcommand{\lyxaddress}[1]{
\par {\raggedright #1
\vspace{1.4em}
\noindent\par}
}

\@ifundefined{definecolor}
 {\usepackage{color}}{}
\@ifundefined{definecolor}{\usepackage{color}}{}
\date{}

\usepackage{babel}

\makeatother

\usepackage{babel}
\begin{document}

\title{Energy distributions for ionization in ion-atom collisions}

\author{A. Amaya-Tapia, A. Antillón }

\maketitle

\lyxaddress{\begin{center}
{\scriptsize{}\vspace{1mm}
 Instituto de Ciencias Físicas, Universidad Nacional Autónoma de México,
AP 48-3, Cuernavaca, 62251 Morelos, México } 
\par\end{center}}
\begin{abstract}
In this paper we discuss how through the process of applying the Fourier
transform to solutions of the Schrödinger equation in the Close Coupling
approach, good results for the ionization differential cross section
in energy for electrons ejected in ion-atom collisions are obtained.
The differential distributions are time dependent and through their
time average, the comparison with experimental and theoretical data
reported in the literature can be made. The procedure is illustrated
with reasonable success in two systems, $p+H$ and $p+He$, and is
expected to be extended without inherent difficulties to more complex
systems. This allows advancing in the understanding of the calculation
of ionization processes in ion-atom collisions. \end{abstract}
\begin{quote}
\medskip{}


\raggedright{\textbf{PACS:} number(s): 34.50.Fa, 34.10.+x} 
\end{quote}

\section{Introduction}

The ejection of electrons involves the greatest energy transfer compared
to the other electronic processes during collisions of ions with atoms
or molecules. When these electrons are sufficiently energetic can
cause further ionization. Studies of these processes are important
for understanding the energy deposition by fast ions moving through
matter, like happens in cancer therapy \cite{Kirchner:2011}. 

The knowledge of differential cross sections has boosted recently
a more detailed picture of the ionization dynamics of simple atomic
systems impinged by ions \cite{Arthanayaka:2015,Xiao:2014} that has
applications in comet and planetary emission modelling \cite{Santos:2013}.

There are excellent reviews regarding the ejected electron spectra
arising from ion-atom collisions by Rudd et al. \cite{Rudd:1985},
Stolterfoht et al. \cite{Stolterfoht:1997} and Ovchinnikov et al.
\cite{Ovchinnikov:2004}. 

Several theoretical approaches have been applied to the $p+H$ and
$p+He$ systems to study the spectrum of the ionized electrons. Usually
Schrödinger equation for one electron is solved and with approximate
models the effects of other electrons are included when more than
one electron is present. We can divide the theoretical methods into
three types, perturbative methods, numerical solutions of the Schrödinger
equation (direct numerical solution of the time-dependent Schrödinger
equation and coupled channel approximations) and classical calculations.
In most of the papers the impact parameter approximation is used.

In relation to methods of interest for this work, among the perturbative
methods is the one of Schulz et al. \cite{Schulz:1996} which uses
analytical initial and final wave functions in the first order of
a distorted wave series and also a Hartree-Fock potential that does
not include the projectile-core interaction. To include post-collision
interactions (PCI) the transition amplitudes are multiplied by a Coulomb
factor that depends on the electron and the projectile velocities.

In the group of direct numerical solutions, Schultz et al. \cite{Schultz:2002}
obtains a direct solution of the time dependent Schrödinger equation
on a configuration space lattice. They solve the Schrödinger equation
going from the coordinate space to momentum space and then back to
configuration space for each time interval (Fourier Collocation Method).

Coupled channel methods usually use the impact parameter approximation
and consist in expanding the wave function, in one or two centers,
on a truncated basis set. The Schrödinger equation is replaced by
systems of differential equations for the coefficients appearing in
the expansion and these equations are solved for large times. Sidky
et al. \cite{Sidky:1998} started with the wave function in a momentum
grid and transformed it to the coordinate space where the system of
differential equations is solved directly. They go back to the momentum
space to analyze the wave function for obtaining the electron ejected
spectrum. Chassid et al. \cite{Chassid:2002} have propagated in time
the wave function by using the Fourier Collocation Method and carried
out the analysis of the ionization wave packet by making histograms
for each point in the configuration space mesh in order to calculate
differential probabilities in emitted electron energies. Fu et al.
\cite{Fu:2001} used pseudo-states only in one center and bound states
in the two centers and solve the Schrödinger equation in coordinate
space.

In classical calculations, the classical equations of motion are solved,
and a quantum momentum distribution for the initial states of the
ionized electrons is used. Schulz et al. \cite{Schulz:1996} perform
classical Monte Carlo calculations that includes the projectile-core
interaction and the PCI factor mentioned above.

In this work, an independent particle approach that makes use of the
two centers Close-Coupling method in the impact parameter approximation
is presented. It applies to simple systems to advance in the understanding
of the ionization processes calculations, by obtaining the energy
distribution of emitted electrons in ion-atom collisions. This is
shown on two simple reactions ($p+H$ and $p+He$) where we presume
our approach is more direct than some previous ones. The continuum
spectrum of the ionized electron is obtained through a Fourier transform
of the ionization amplitude given in the coordinate space. The found
differential sections are time dependent, so it is convenient to make
a time average to compare with experimental data since the time dependency
at large distances does not decrease asymptotically because comes
from the phase appearing in the basis functions.

We think that this process is more transparent than other schemes
where is necessary to do an interpolation of a large set of ionization
probability amplitudes in order that the time dependency they have,
could appear as a global phase in the sum of the different pseudo-states
amplitudes and therefore can be removed \cite{Reading:2005}.

\section{Method}

\ 

The Close Coupling method used in this work is a time-dependent impact-parameter
two-center approach \cite{Lin:1991}. In this model, the nuclei follow
classical trajectories and the electron dynamics is described by the
Schrödinger equation. For the $p+H$ system the interactions between
the three particles are Coulomb interactions. In the $p+He$ system
the potential model 
\begin{equation}
V\left(r_{B}\right)=\left(-\frac{1}{r_{B}}+\frac{\left(-1-0.4042\, r_{B}\right)}{r_{B}}e^{-1.9394\, r_{B}}\right)\label{eq:pot}
\end{equation}
substitutes the Coulomb interactions for the electron-$He^{+}$ system.
Here $\mathbf{r}_{B}\left(\mathbf{r}_{A}\right)$ is the electron
coordinate with respect to center B(A). The time dependence is given
through the internuclear distance, $\mathbf{R}=\mathbf{r}_{B}-\mathbf{r}_{A}=\mathbf{b}+Vt\hat{k}$
, where $V$ is the projectile velocity with respect to the target
center, $\mathbf{b}=\left(b,\gamma\right)$ the impact parameter vector
and $t$ the time. The Schrödinger equation is solved along straight
lines for each impact parameter $b$. 

The wave function is expanded in atomic functions $\phi_{nlm}\left(\mathbf{r},t\right)$
as 
\begin{equation}
\Psi\left(\mathbf{r}_{B};\mathbf{b},t\right)={\displaystyle \sum_{n,l,m}}a_{nlm}^{B}\left(\mathbf{b},t\right)\phi_{nlm}^{B}\left(\mathbf{r}_{B},t\right)+e^{i\left(\mathbf{V}\cdot\mathbf{r}_{B}-\frac{1}{2}V^{2}t\right)}{\displaystyle \sum_{n,l,m}}a_{nlm}^{A}\left(\mathbf{b},t\right)\phi_{nlm}^{A}\left(\mathbf{r}_{A},t\right).\label{eq:wfc}
\end{equation}

The phase factor in front of the projectile centered expansion takes
account of the projectile translational motion. The atomic basis are
expanded in even-tempered functions, useful in studying different
collisional processes, 
\begin{equation}
\phi_{nlm}^{j}\left(\mathbf{r}_{j},t\right)=e^{-i\epsilon_{nl}^{j}t}{\displaystyle Y_{lm}\left(\theta_{j},\phi_{j}\right)r_{j}^{l}\sum_{k=1}^{k_{max}}s_{nl}^{jk}\left[\left(2\alpha_{l}^{j}\beta_{l}^{jk}\right)^{2l+3}/\left(2l+2\right)!\right]^{1/2}e^{-\alpha_{l}^{j}\beta_{l}^{jk}r_{j}}}.\label{eq:atfc}
\end{equation}

The parameters \{$\alpha,\beta,k$\} are chosen by fitting the atomic
energy levels of the isolated atomic systems when their Hamiltonians
are diagonalized. This procedure determines the coefficients $s_{nl}^{jk}$.
The square root factor is the normalization constant for the radial
component and $Y_{lm}\left(\theta_{j},\phi_{j}\right)$ are the normalized
spherical harmonics. The diagonalization procedure yields both wave
functions of negative energy and states of positive energy commonly
called pseudostates. The amplitudes $a_{nlm}^{j}\left(\mathbf{b},t\right)$
in Eq.\ \ref{eq:wfc} are numerically obtained by solving the set
of coupled differential equations that results from the projection
of the Schrödinger equation onto the atomic functions, with the electron
located initially in the target ground state.

The coordinate space solutions only allow a discrete energy distribution
for the ionized electrons. This leads us to the non-trivial point
of how to derive the differential cross section in electronic energy
\cite{Reading:1979}. With the momentum space representation is possible
to get continuous spectra, an option that we address in the next paragraphs.

The Fourier transform of the spatial atomic functions is given by
\begin{equation}
\begin{array}{cc}
\tilde{\phi}_{nlm}^{j}\left(\mathbf{p}_{j},t\right) & =\left(2\pi\right)^{-3/2}\int d\mathbf{{\normalcolor r}}_{j}\, e^{i\mathbf{r}_{j}\cdot\mathbf{p}_{j}}\phi_{nlm}^{j}\left(\mathbf{{\normalcolor r}}_{j},t\right),\end{array}\label{eq:ftr}
\end{equation}
that can be written as the sum (see Appendix)

\begin{equation}
\tilde{\phi}_{nlm}^{j}\left(\mathbf{p}_{j},t\right)=e^{-i\epsilon_{nl}^{j}t}\sum_{k=1}^{k_{max}}s_{nl}^{jk}F_{l}^{jk}\left(p_{j}\right)Y_{lm}\left(\vartheta_{j},\varphi_{j}\right),\label{eq:cm-1}
\end{equation}
 where 
\begin{equation}
F_{l}^{jk}\left(p_{j}\right)=\frac{-\left(-\mathfrak{i}\right)^{l}2^{2l+3}\left(\alpha_{l}^{j}\beta_{l}^{jk}\right)^{l+5/2}\left(l+1\right)!}{\left[\pi\left(2l+2\right)!\right]^{1/2}}\frac{\left(p_{j}\right)^{l}}{\left(\left[\alpha_{l}^{j}\beta_{l}^{jk}\right]^{2}+\left[p_{j}\right]^{2}\right)^{l+2}}.\label{eq:rad}
\end{equation}

For an impact parameter $\mathbf{b}$, the probability density that the
electron can be found in the asymptotic state, with spherical momentum
components $\left(p_{B},\vartheta_{B},\varphi_{B}\right)$ is calculated
as \cite{Chassid:2002}

\begin{equation}
\begin{array}{l}
\left|\tilde{\Psi}\left(\mathbf{p}_{B};\mathbf{b},t\right)\right|^{2}\equiv\left|\left(2\pi\right)^{-3/2}\int d\mathbf{r}_{B}\, e^{-i\mathbf{p}_{B}\cdot\mathbf{r}_{B}}\Psi\left(\mathbf{r}_{B};\mathbf{b},t\right)\right|^{2}\\
\\
=\left|{\displaystyle \sum_{n,l,m}}a_{nlm}^{B}\left(\mathbf{b}\right)\tilde{\phi}_{nlm}^{B}\left(\mathbf{p}_{B},t\right)+e^{-i\left(\mathbf{p}_{B}\cdot\mathbf{R}-\frac{1}{2}V^{2}t\right)}{\displaystyle \sum_{n,l,m}}a_{nlm}^{A}\left(\mathbf{b}\right)\tilde{\phi}_{nlm}^{A}\left(\mathbf{p}_{A},t\right)\right|^{2},
\end{array}\label{eq:pr}
\end{equation}
 where $a_{nlm}^{j}\left(\mathbf{b}\right)=a_{nlm}^{j}\left(\mathbf{b},t_{large}\right)$.
For $t=t_{large}$ the ionization channel $\tilde{\Psi}_{I}$ is represented
by difference between the total wave function and its projection onto
the bound channels $\tilde{\Psi}_{L}$ \cite{Chassid:2002},
\begin{equation}
\begin{array}{cl}
\tilde{\Psi}_{I}\left(\mathbf{p}_{B};\mathbf{b},t\right) & =\tilde{\Psi}\left(\mathbf{p}_{B};\mathbf{b},t\right)-\tilde{\Psi}_{L}\left(\mathbf{p}_{B};\mathbf{b},t\right)\end{array}\label{eq:iw}
\end{equation}
where
\begin{equation}
\tilde{\Psi}_{L}\left(\mathbf{p}_{B};\mathbf{b},t\right)={\displaystyle \sum_{\begin{array}{c}
n,l,m,j\\
\epsilon_{n}^{j}<0
\end{array}}}\left(\int d\mathbf{p}_{B}'e^{i\zeta\delta_{j,A}}\phi_{nlm}^{j}\left(\mathbf{p}_{B}',t\right)\tilde{\Psi}\left(\mathbf{p}_{B}';\mathbf{b},t\right)\right)e^{i\zeta\delta_{j,A}}\phi_{nlm}^{j}\left(\mathbf{p}_{B},t\right),\label{eq:lw}
\end{equation}

and $\zeta=\mathbf{V}\cdot\mathbf{r}_{B}-\frac{1}{2}V^{2}t$. We point out
that the probability calculated with the function $\tilde{\Psi_{I}}\left(\mathbf{p}_{B};\mathbf{b},t\right)$
includes the interference terms between the two centers. 

To include the effect of the second electron on the ionization probability
in the case of $He$ as target, we use the common procedure referenced
in the literature as the binomial combination of probabilities of
isolated particles \cite{Spranger:2004}, allowing the second electron
($\left\{ 2\right\} $) to end in any bound state

\begin{equation}
P_{I}\left(\mathbf{p}_{B};\mathbf{b},t\right)=2\left|\tilde{\Psi}_{I}\left(\mathbf{p}_{B}\left\{ 1\right\} ;\mathbf{b},t\right)\right|^{2}\left(1-\int d\mathbf{p}_{B}\left\{ 2\right\} \, P_{I}\left(\mathbf{p}_{B}\left\{ 2\right\} ;\mathbf{b},t\right)\right).\label{eq:wfin}
\end{equation}

The angular and momentum distributions $D\left(p_{B},\vartheta_{B}\right)d\vartheta_{B}dp_{B}$
of the ionized electrons are obtained by integrating the ionization
probability density over $\varphi_{B}$ and\textcolor{red}{{} }$b$,
\begin{equation}
D\left(p_{B},\vartheta_{B}\right)d\vartheta_{B}dp_{B}=d\vartheta_{B}dp_{B}\int b\, db\, d\gamma\int d\varphi_{B}\left|\tilde{\Psi}_{I}\left(\mathbf{p}_{B};\mathbf{b},t\right)\right|^{2}.\label{eq:d}
\end{equation}

One purpose in this work is to extract the distribution in energy.
Then by making the appropriate change of variables between the differentials
in momentum $p_{B}$ and energy $E_{B}$ the distribution function
can be written as 
\begin{equation}
D\left(p_{B},\vartheta_{B}\right)d\vartheta_{B}dp_{B}=\frac{1}{\sqrt{2E_{B}}}D\left(p_{B}\left[E_{B}\right],\vartheta_{B}\right)d\vartheta_{B}dE_{B}.\label{eq:de}
\end{equation}

The calculation of this distribution in the target frame takes into
account a frame transformation applied to the functions centered at
the projectile. 

So far we have been focused on the analytical determination of the
distribution functions, but we have not completely specified the atomic
functions. Next, we briefly describe how this has been done. The parameters
$\alpha,\beta$ and $k$ appearing in Eq. \ref{eq:atfc} were determined
by fitting the low energy levels of the hydrogen or helium atoms to
the experimental values. For this optimization procedure the subroutine
Minuit from the CERN library was used and the values obtained are
displayed in Table~\ref{Ref:table:parameters}. Once the atomic functions
had been obtained, the system of time-dependent equations for the
amplitudes $a_{nlm}^{i}\mathbf{\left(b,t\right)}$ were numerically solved
over 232 trajectories, chosen randomly in the impact parameter range
from 0.02 to 25.0 a.u. The dynamics was followed for internuclear
distances larger than of 100 a.u., distance at which the value of
the amplitudes had already converged. The calculations were made for
three projectile energies, 25, 50 and 100 keV in the case of $H$
target and for 50 keV in the $He$ target.

\noindent 
\begin{table}[b]
\centering{}%
\begin{tabular}{|c|c|c|c|}
\hline 
\multicolumn{4}{|c|}{$H$}\tabularnewline
\hline 
$l$ & $\alpha_{l}$ & $\beta_{l}$  & $k_{max}$\tabularnewline
\hline 
s  & 3.206$\times$10$^{-2}$  & 1.196  & 15 \tabularnewline
\hline 
p  & 4.868 $\times$10$^{-2}$  & 1.150  & 15 \tabularnewline
\hline 
d  & 5.892 $\times$10$^{-2}$  & 1.249  & 30 \tabularnewline
\hline 
\end{tabular}%
\begin{tabular}{c|c|c|}
\hline 
\multicolumn{3}{c|}{$He$}\tabularnewline
\hline 
$\alpha_{l}$ & $\beta_{l}$  & $k_{max}$\tabularnewline
\hline 
5.051$\times$10$^{-2}$  & 1.108 & 10\tabularnewline
\hline 
7.114$\times$10$^{-2}$  & 1.187  & 15 \tabularnewline
\hline 
9.576 $\times$10$^{-2}$  & 1.130  & 15 \tabularnewline
\hline 
\end{tabular}\caption{Even-tempered function parameters, used in Eq. \ref{eq:atfc} for
the $H$ and $He$ cases.}
\label{Ref:table:parameters}
\end{table}

\section{Basis set comparative analysis}

\ For 50 keV protons colliding with hydrogen and an internuclear
distance of $Vt=100$ a. u., an assessment of the reliability of our
basis set is done in Table \ref{table:2} by comparing some of the
cross section values here obtained for electron transfer, excitation
and ionization, with those of Winter \cite{Winter:2009} and few of
them with Fitzpatrick et al\textsl{. }\cite{Fitzpatrick:2007}. The
former used an (s,p,d) basis with 88 states at each center and the
latter used an (s-g) basis with 22 states per $l$ and $m$, and both
use a similar approach than that used here. We note in Table~\ref{table:2}
that our calculations reproduce with good accuracy both results, even
in the partial cross sections, which are more sensitive to the model
details than the total cross section. A comparison with experimental
data for the total electron transfer cross section shows that our
value at 50 keV differs by 10 \% from that of McLure et al. \cite{McLure:1966}
($1.10\times10^{-16}\, cm^{2}$ $\pm5\%$) measured at 48 keV.

Concerning the relation between theoretical and experimental ionization
cross sections \cite{Chassid:2002}, numerical calculations are consistently
higher than experimental data in the (50-60)-keV range. For example,
Kerby et al. \cite{Kerby:1995} report the value $1.44\times10^{-16}\, cm^{2}$
at 48 keV in the projectile energy, which is about 18 \% lower than
the cross section reported in this work, and 6 \% higher than $1.36\times10^{-16}\, cm^{2}$,
suggested by Rudd et al. \cite{Rudd:1985} after the analysis of several
experiments.

Toshima \cite{Toshima:1999} applied a similar close-coupling formalism
to proton colliding on atomic hydrogen and made a careful analysis
of the convergence of the pseudostate representation, in basis size
and in the range of angular momentum quantum numbers needed to be
taken into account at each center. According to him, the size of a
symmetric basis set (same number of functions in each center) is smaller
for achieving convergence, than the size in the asymmetric case. A
comparative analysis for the ionization probabilities as function
of the impact parameter are shown in Fig.\ 1 for different basis
sizes \cite{Amaya:2012}. The convergence trend of our calculations
can be followed as the size of the basis increase. For reference purpose
we include the results of Toshima corresponding to a basis with 197
states in each center, and quantum numbers $l$ up to 5. It can be
seen that our curves get closer to the Toshima curves at lower and
higher impact parameters. Around the peak, our values differ from
those of Toshima by about 11 \%. Considering these results, we found
that a good basis for the purpose of this work include 13 s's, 13
p's and 14 d's, given in total 81 states centered at each proton,
41 of them correspond to pseudostates. For $He$ we include, 9 s's,
11 p's and 10 d's, given in total 61 states, 41 of them correspond
to pseudostates. All the computed energy levels are shown in Tables~\ref{table:3}
and~\ref{table:4}. 
\begin{figure}
\centering{}\includegraphics[scale=1.1]{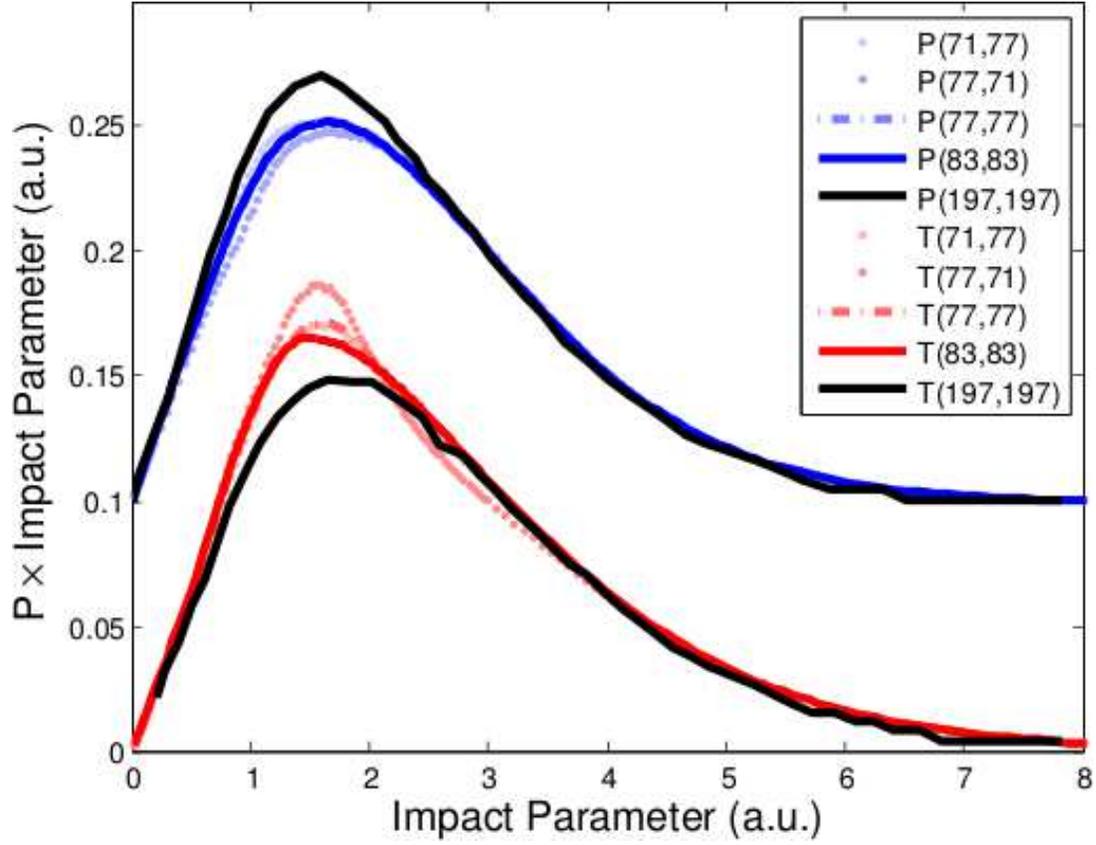}\caption{The ionization probabilities as a function of the impact parameter
for the $p+H$ system at 50 keV are shown. This figure displays their
convergence process as a function of basis sizes. In the legends box
we introduce the notation T(x,y) and P(x,y) that is associated to
the contributions to the ionization probabilities from the continuum
states centered at the target and at the projectile respectively,
where x and y are the considered basis sizes in target and projectile,
respectively. The curves labeled P are shifted up 0.1 a.u. for making
clearer the convergence process. Curves in black are results of Ref.\ \cite{Toshima:1999}.}
\end{figure}

\begin{table}
\centering{}%
\begin{tabular}{|c|c|c|c|c|c|c|c|c|}
\cline{3-9} 
\multicolumn{1}{c}{} &  & 1s  & 2s  & 3s  & 2p  & 3p  & 3d  & TC\tabularnewline
\hline 
\multirow{3}{*}{ET} &  {*} & 0.696 & 0.141  & 0.042  & 0.038  & 0.012  & 0.001  & 0.99\tabularnewline
\cline{2-9} 
 & W & 0.698 & 0.141 & 0.043 & 0.037 & 0.012 & 0.001 & 1.00\tabularnewline
\cline{2-9} 
 & F  & 0.695  & 0.144  &  & 0.048  &  &  & \tabularnewline
\hline 
\multirow{2}{*}{EX} & {*} &  & 0.16 & 0.04 & 0.72 & 0.13 & 0.04 & 1.27\tabularnewline
\cline{2-9} 
 & W  &  & 0.17  & 0.03  & 0.75  & 0.13  & 0.03  & 1.30\tabularnewline
\hline 
\multirow{2}{*}{EI} & {*} & \multicolumn{6}{c|}{} & 1.66\tabularnewline
\cline{2-9} 
 & W & \multicolumn{6}{c|}{} & 1.73\tabularnewline
\hline 
\end{tabular}\caption{Cross sections (in units of 10$^{-16}$cm$^{2}$) in $p+H$ collisions
at 50 keV and $Vt=100$ a. u. for different processes: electron transfer
(ET), direct excitation (EX) to $H$ states, ionization (EI) and total
cross sections (TC). The results presented for the cross sections
are from Winter's (W) \cite{Winter:2009}, Fitzpatrick's et al. (F)
\cite{Fitzpatrick:2007} and this work ({*}). }
\label{table:2}
\end{table}

\begin{table}
\centering{}%
\begin{tabular}{|c|c|c|c|}
\cline{2-4} 
\multicolumn{1}{c|}{} & $l=0$  & $l=1$  & $l=2$\tabularnewline
\hline 
$E_{1l}$  & -0.50000  &  & \tabularnewline
\hline 
$E_{2l}$  & -0.12500  & -0.12500  & \tabularnewline
\hline 
$E_{3l}$  & -0.05556  & -0.05556  & -0.05556\tabularnewline
\hline 
$E_{4l}$  & -0.03125  & -0.03125  & -0.03125\tabularnewline
\hline 
$E_{5l}$  & -0.01996  & -0.02000  & -0.02000\tabularnewline
\hline 
$E_{6l}$  & -0.01374  & -0.01387  & -0.01389\tabularnewline
\hline 
$E_{7l}$  & -0.00994  & -0.01014  & -0.01018\tabularnewline
\hline 
$E_{8l}$  & -0.00719  & -0.00760  & -0.00773\tabularnewline
\hline 
$E_{9l}$  & 0.01956  & 0.00396  & 0.00000\tabularnewline
\hline 
$E_{10l}$  & 0.14791  & 0.05674  & 0.02918\tabularnewline
\hline 
$E_{11l}$  & 0.53874  & 0.19935  & 0.09865\tabularnewline
\hline 
$E_{12l}$  & 1.63458  & 0.54938  & 0.24404\tabularnewline
\hline 
$E_{13l}$  & 4.80275  & 1.42889  & 0.53182\tabularnewline
\hline 
$E_{14l}$  &  & 3.97152  & 1.08438\tabularnewline
\hline 
$E_{15l}$  &  &  & 2.12542\tabularnewline
\hline 
$E_{16l}$  &  &  & 4.06261\tabularnewline
\hline 
\end{tabular}\caption{The eigenenergies (in a.u.) for $H$ are reported in this table. They
are computed by diagonalizing the corresponding Hamiltonian.}
\label{table:3}
\end{table}

\begin{table}
\centering{}%
\begin{tabular}{|c|c|c|c|}
\cline{2-4} 
\multicolumn{1}{c|}{} & $l=0$  & $l=1$  & $l=2$\tabularnewline
\hline 
$E_{1l}$  & -0.81685  &  & \tabularnewline
\hline 
$E_{2l}$  & -0.15324  & -0.12670  & \tabularnewline
\hline 
$E_{3l}$  & -0.06335  & -0.05613  & -0.05556\tabularnewline
\hline 
$E_{4l}$  & -0.03447  & -0.03150  & -0.03125\tabularnewline
\hline 
$E_{5l}$  & -0.02161  & -0.02011  & -0.02000\tabularnewline
\hline 
$E_{6l}$  & -0.01470  & -0.01373  & -0.01387\tabularnewline
\hline 
$E_{7l}$  & -0.01026  & -0.00594  & -0.00984\tabularnewline
\hline 
$E_{8l}$  & 0.03274  & 0.04105  & 0.00140\tabularnewline
\hline 
$E_{9l}$  & 0.40389  & 0.18039  & 0.03463\tabularnewline
\hline 
$E_{10l}$  &  & 0.51979  & 0.11004\tabularnewline
\hline 
$E_{11l}$  &  & 1.29732  & 0.26716\tabularnewline
\hline 
$E_{12l}$  &  & 3.06058  & 0.58697\tabularnewline
\hline 
\end{tabular}\caption{The eigenenergies (in a.u.) for $He$ are reported in this table.
They are computed by diagonalizing the one electron Hamiltonian with
the model potential of Eq. \ref{eq:pot}.}
\label{table:4}
\end{table}

\section{Results and discussion}

It is known that the pseudostates are not really continuum states
because they decay asymptotically at large distances. However their
use is appropriate to describe ionization. In this sense it has been
shown \cite{Toshima:1999,Sidky:1999} that all the dynamics of the
ionization mainly occurs in a confined spatial region where the interaction
potential has its dominant role and where the ejected-electron distribution
in energy has been already established. Also Lee et al. \cite{Lee:2007}
studied the time evolution of the spectrum of ionized electrons in
the $p+H$ collision and found that in the distance range of 500-5000
a. u. the profiles are very similar but at a distance of 50 a. u.
the distribution profile is still in noticeable evolution. This is
more clear in the analysis of the Fig.2.

In Fig. 2 it is shown four energy distribution curves of ionized electrons
in the $p+H$ collision, two of which are numerically equal. The orange
curve shows the result obtained when the final ionization function
is evaluated at a distance $Vt=200$ a. u. and the green circles represent
the distribution when the ionization function is made up with coefficients
$a_{nlm}^{j}$ evaluated at $Vt=110$ a. u., multiplied by the atomic
functions with their time depending phases evaluated at the time corresponding
to $Vt=200$ a. u. We see that there is a complete agreement among
those curves, meaning that the evolution was due exclusively to the
contribution of the time dependent phases and not to the interactions
present in the Hamiltonian through the potential, and also that the
electronic density has almost completed its evolution to a final state.

The other pair of curves represents a similar analysis, but it is
done at a shorter distance $Vt$. The black curve shows the result
obtained when the final ionization function is evaluated at a distance
$Vt=50$ a.u. while the blue curve represents the distribution when
the ionization function is formed with coefficients $a_{nlm}^{j}$
evaluated at $Vt=110$ a.u., multiplied by the atomic functions with
their time depending phases evaluated at the time corresponding to
$Vt=50$ a.u. In this case we clearly note a pronounced difference
among both distributions meaning that the electron density is still
changing due to the potential interaction and therefore the final
distance at which the calculatons were made has to increase to guarantee
a negligible influence of the potential. 

The curves displayed in the inset of Fig. 2 are some energy distributions
of ionized electrons in the $p+H$ collision calculated with the ionization
function evaluated at different distances $Vt$, where the potential
interaction does not affect the electronic distribution and then the
variation of those distributions are due essentially to the time dependent
phases.

Finally, the small structure known as ECC shown at the matching region
around 27 eV, where the velocity of ejected electrons and projectile
are similar, appears clearly at each distribution for a given $Vt$.

\begin{figure}
\centering{}\includegraphics[scale=0.8]{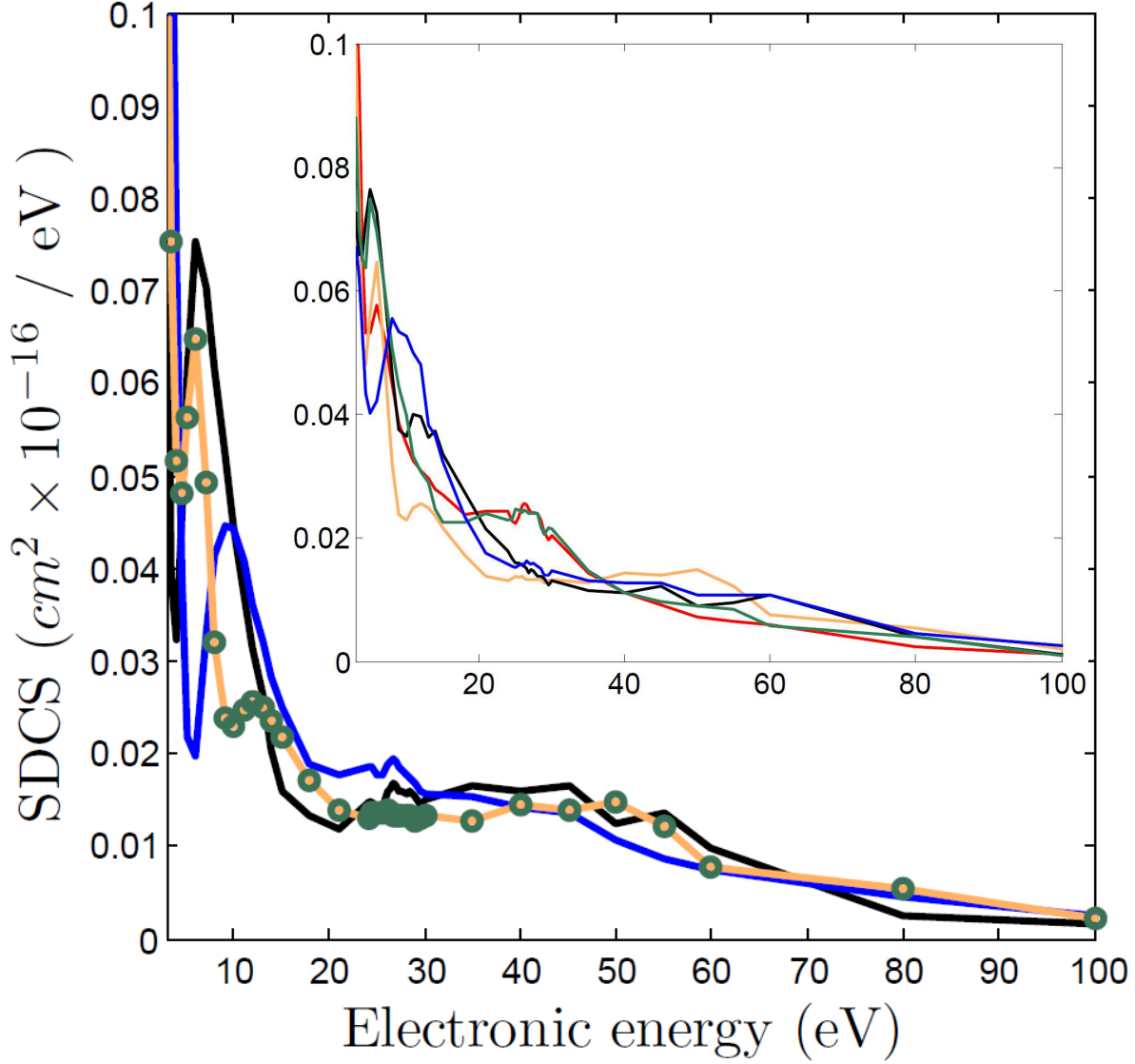}\caption{Single differential cross sections for the $p+H$ system as a function
of the emitted electron energy, calculated at the different distances
$Vt$ (in a. u.). The orange curve and the one described by green
circles represent the energy distribution at $Vt=200$ a. u. calculated
as described in the text, while the blue and black curves have a similar
meaning for $Vt=50$ a. u. In the inset there are some energy distribution
curves that sample the region occupied by the full set of distributions
when the time dependent phase is varied through the distance $Vt$.
The different colors red, orange, black, green and blue, correspond
to $Vt$ of 100, 200, 300, 400 and 500 a. u.}
\end{figure}

In our study the distance interval $V\Delta t\sim500$ a. u. corresponds
to a time interval $\Delta t\sim10^{-15}$ seconds that is close to
the inverse of the difference between the lower pseudo-states energies.
On the other hand, the data acquisition time in experimental runs
is much more larger than this estimated time and for this reason the
oscillations shown in Fig. 2 would not be seen in the experimental
results. In order to compare with the experimental measurements, a
key step in the present work is to make a temporal average of the
calculated distributions for several $Vt$. This idea is discussed
in the analysis of Fig. 3.

In Fig. 3 the black line shows results for the electron energy differential
cross section for the $p+H$ system. That was calculated taking the
time average of 40 distributions at different $Vt$ in the range of
100 to 550 a. u. In magenta circles is exhibited the experimental
measurements obtained by Kerby et al. \cite{Kerby:1995}, whose uncertainties
in their cross section are of 22 \% above 10 eV, increasing to 26
\% at 1.5 eV and to 50 \% or more at the highest energies in the plot.
What we have noticed in our calculation is that as we include more
distributions in our time average, the mean distribution shows a smoother
curve that approaches closer to the Kerby et al. experimental profile
and the oscillatory character is pushed into lower energies. At energies
above 20 eV, the differential cross section is larger than that shown
experimentally and below 20 eV is lower than that. 

Fig. 3 also shows other theoretical calculations using the approximation
of the impact parameter. The results from Chassid et al. \cite{Chassid:2002}
(in blue line) fit well with the experimental data for high energies
but has a significant disagreement at low energies. When our results
are compared with those calculations, there is a significant difference
in the shape of electron-energy differential cross sections. As we
know from the introduction, both methods are similar in the main physical
approximations and this is reflected in the good agreement in the
ionization probability as a function of the impact parameter, as shown
in the inset of Fig. 3. However the difference in the methods arises
at the level of spectrum extraction as is pointed out in the introduction.
One of these differences is that our calculation was done as an averaged
distribution of several $Vt$'s in a range whose values are larger
than the single $Vt$ value used by Chassid et al. where there is
an important perturbation of the potential. Another difference is
that we used the final wave function in the momentum space for obtaining
the electron energy spectrum, whereas they got the energy spectrum
by using eigenenergies of pseudostates in configuration space.

In the averaged distribution, the ECC structure diminishes considerably
compared to that appearing in single distributions, as shown in Fig.
2. This little structure does not appear in the other theoretical
calculations nor in the experimental data that indeed have not been
measured at small angles and therefore it should not be there. A possible
cause for missing the ECC peak in the work of Chassid et al. \cite{Chassid:2002}
could be that the $Vt$ distance used may not be sufficient to have
the final electronic density profile for building the structure.The
data calculated by Fu et al. \cite{Fu:2001}have a good agreement
with experimental data but the ECC structure does not appear because
a single center for pseudostates was considered.

\begin{figure}
\centering{} \includegraphics[scale=0.8]{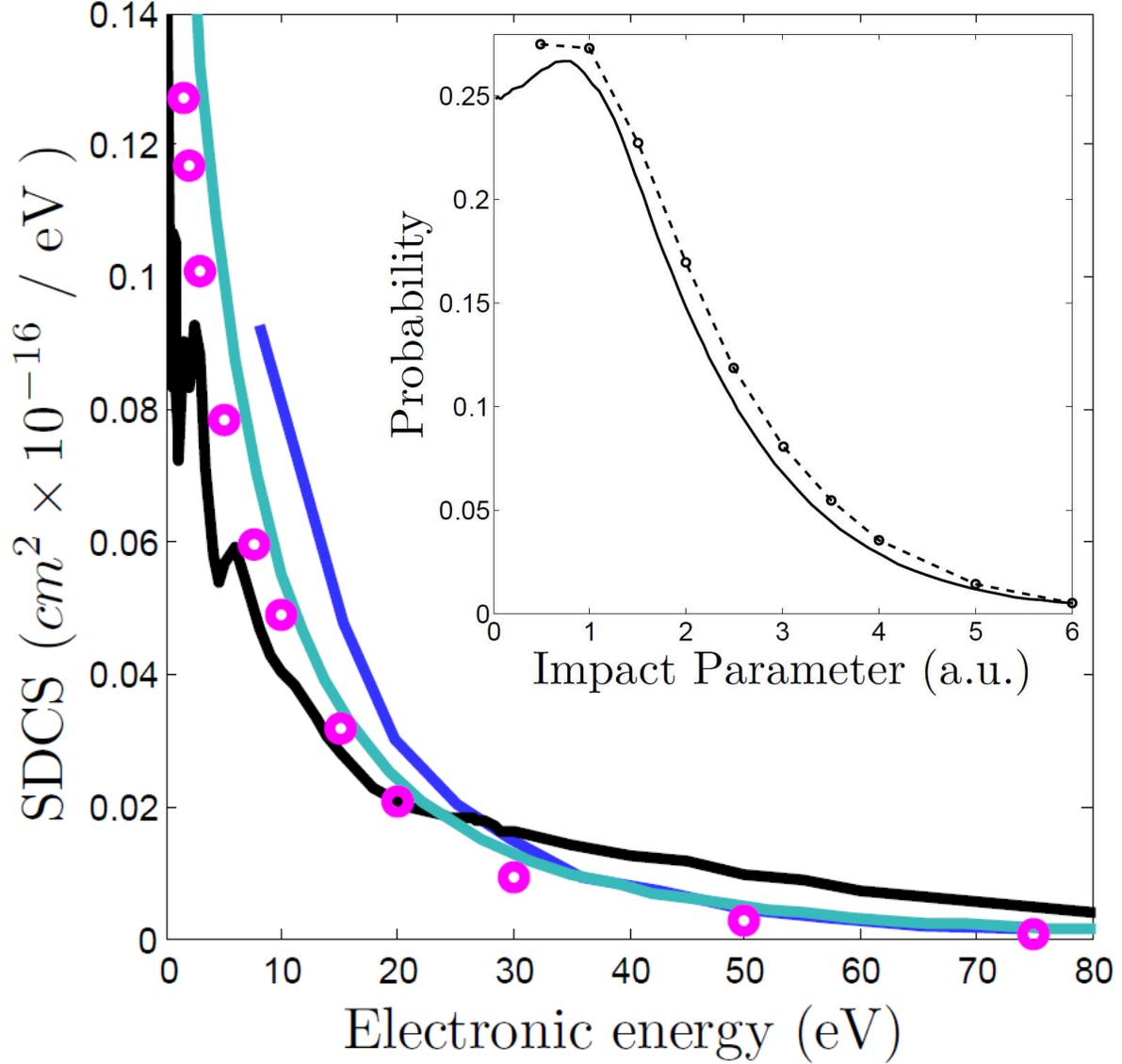}\caption{Ionization distribution in energy for the $p+H$ system at 50 keV.
Black line shows the average distribution on time calculated in this
work. Circles are the experimental data from Ref. \cite{Kerby:1995},
the turquoise line shows calculations from Ref. \cite{Fu:2001} and
the blue line is theoretical data taken from Ref. \cite{Chassid:2002}.
In the inset, a comparison between our results and those of Chassid
et al. \cite{Chassid:2002} for the ionization probability as function
of the impact parameter is shown.}
\end{figure}

The ionization cross section is presented in Fig. 4 for different
$Vt$ distances in the region from 100 to 550 a. u. We can notice
the oscillatory character of its behavior that arises from the time
dependent phases that multiply the stationary atomic functions (see
Eqs. \ref{eq:wfc}, \ref{eq:atfc}). The variation of the ionization cross section is roughly
in the $1.6-1.75\times10^{-16}$ $cm^{2}$ range and the time average
of those values gives $1.68\times10^{-16}$ $cm^{2}$. Some authors
(\cite{Chassid:2002,Sidky:1999}) have identified this behavior and
have suggested a time average procedure but have assumed that the
average should be very close to their found solutions and have not
proceeded to do further calculations.

\begin{figure}

\begin{centering}
\includegraphics[angle=-90,scale=0.6]{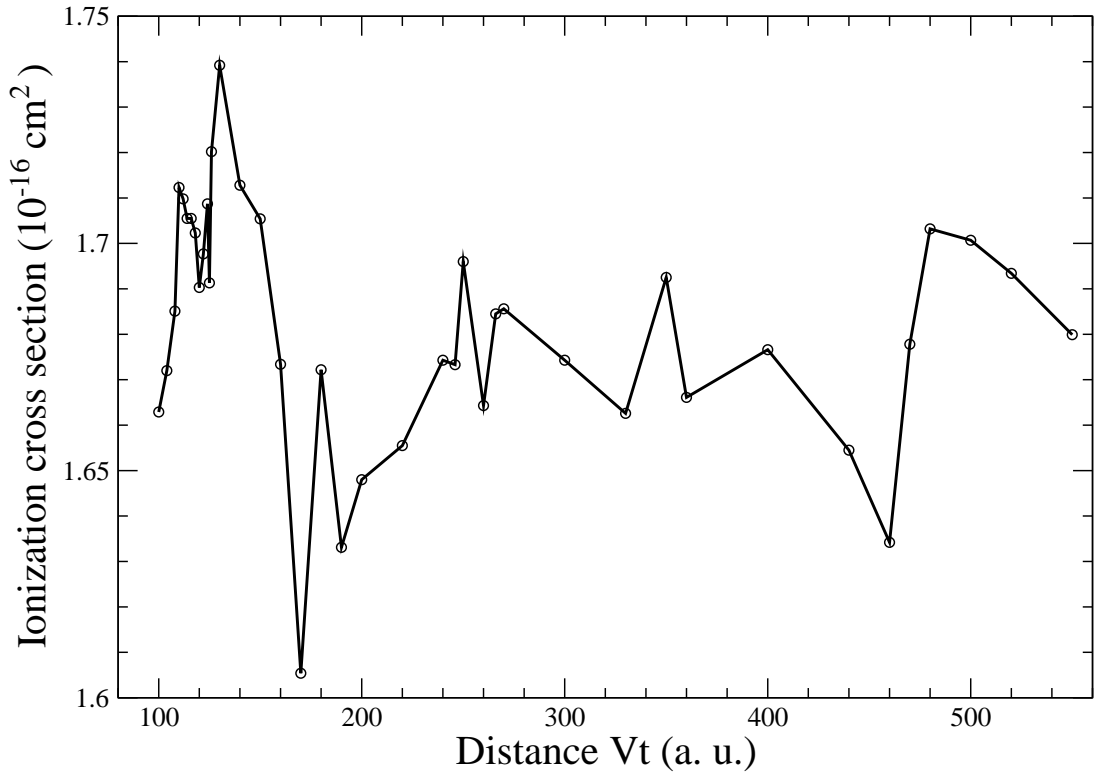}\caption{Ionization cross section for the $p+H$ system as a function of the
distance $Vt$. The oscillations are due to the time dependent phases
appearing in the atomic wave functions.}

\par\end{centering}

\end{figure}

In Fig. 5 the electron-energy differential cross sections calculated
at three incidents energies of 25, 50 and 100 keV and at single distance
of $Vt=100$ a. u. are shown. The full lines represent calculations
made with the differential probability in energy that includes the
terms of interference between the two centers. The construction of
the differential probability sometimes is done subtracting from the
total wave function the projection of each set of functions, anchored
at their respective center, on their own set of bound wave functions.
The dashed lines represent calculations with probabilities obtained
under this scheme. Both calculations differ by a neglectable amount
that is not important when the time average is performed to make a
comparison with experimental data, since this difference is much smaller
than the distribution changes when several $Vt$ are used. Also the
overlap has the tendency to diminish with a larger $Vt$. The main
features of the distribution calculated at a projectile energy of
50 keV are also found at the other energies of 25 and 100 keV. Similar
results are also found in $p+He$ system but are not shown in the
manuscript.

\begin{figure}
\centering{}\includegraphics[angle=-90,scale=0.6]{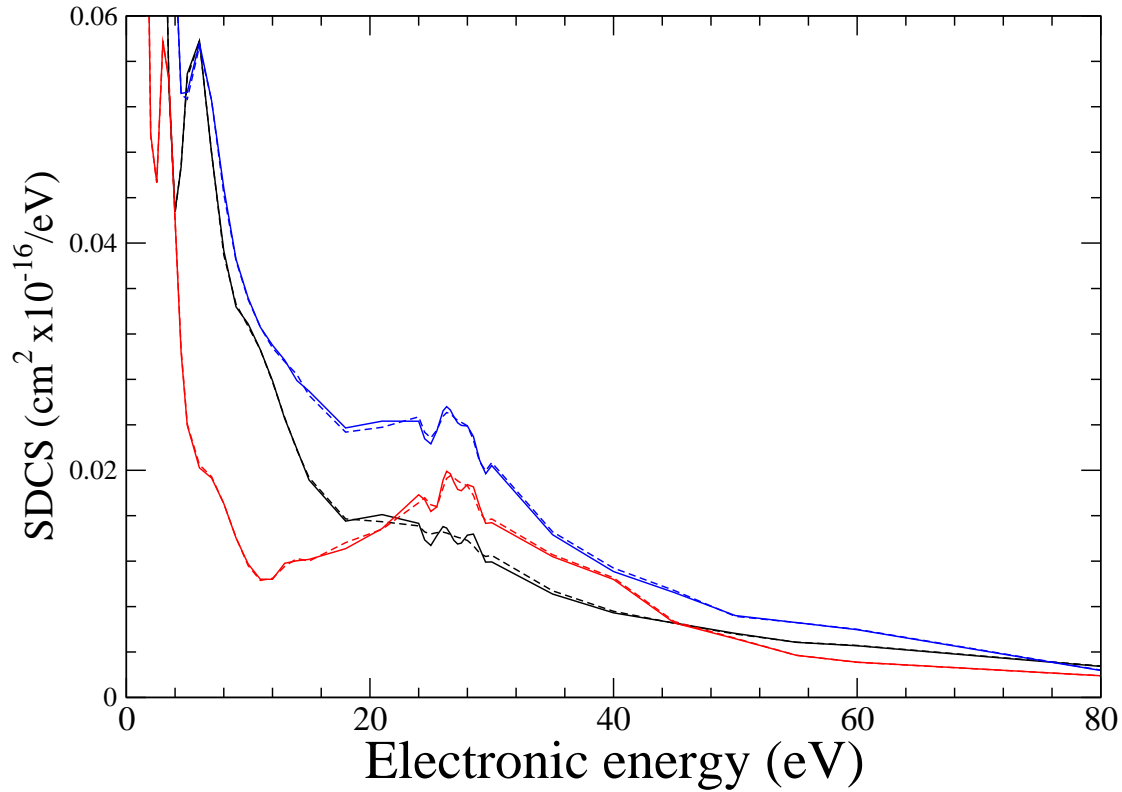}\caption{Ionization distribution in energy for the $p+H$ system at projectile
energies of 25, 50 and 100 keV, at $Vt=100$ a. u., are represented
in red, blue and black colors respectively. Full lines represent our
results calculated with the coherent sum of the wave functions, and
dashed lines, when overlaps between the two sets of wave functions
anchored in each center are not included. We notice that the overlap
is not important and therefore can be ignored.}
\end{figure}

The curves displayed in Fig. 6 are some energy distributions of ionized
electrons in the $p+He$ collision calculated with the ionization
function evaluated at different distances $Vt$, where the potential
interaction does not affect the electronic distribution and then the
variation of those distributions are due essentially to the time dependent
phases. Similarly to hydrogen case, a more notorious structure is
seen in the cusp electrons region for each calculated curve at a given
$Vt$. After the temporal average is done, the structure maintains
its presence there, as is shown in Fig. 7.

In Fig. 7, the single differential cross section as function of the
energy of ejected electrons obtained in this work is compared with
several experimental data and theoretical calculations reported in
the literature for the $p+He$ system. We understand that experimental
data have been normalized following different schemes and therefore
is meaningful to compare only their distribution profiles with the
one we got in this work.

For energies larger than 15 eV our calculation follows a path close
to the experimental results, and below that energy there is a slope
change in experimental data measured by Gibson \cite{Gibson:1986}
and Schulz \cite{Schulz:1996}, which is also seen in this work. In
the data of reference \cite{Cheng:1989} does not appear the structure
close to 27 eV since the measurements were done at electron emission
angles above $10\,^{\circ}$. Also, the theoretical curve reported
by Gulyas (Ref. \cite{Schulz:1996}) follows very close the measurements
made by Cheng et al. For the energy range shown in Fig. 7, similarly
to $p+H$, in the $p+He$ system the curve gets a smoother profile
at large energies and the oscillations diminish at lower energies,
as the number of distribution curves increases in the average calculation.

It is noted that in the cases of $H$ and $He$, the best theoretical
agreement with experiment is achieved with various techniques \cite{Fu:2001,Chassid:2002}
and \cite{Schulz:1996} respectively. In our case, the technique that
is applied to the two reactions provides an acceptable agreement with
experiments of both systems.

\begin{figure}
\includegraphics[angle=-90,scale=0.6]{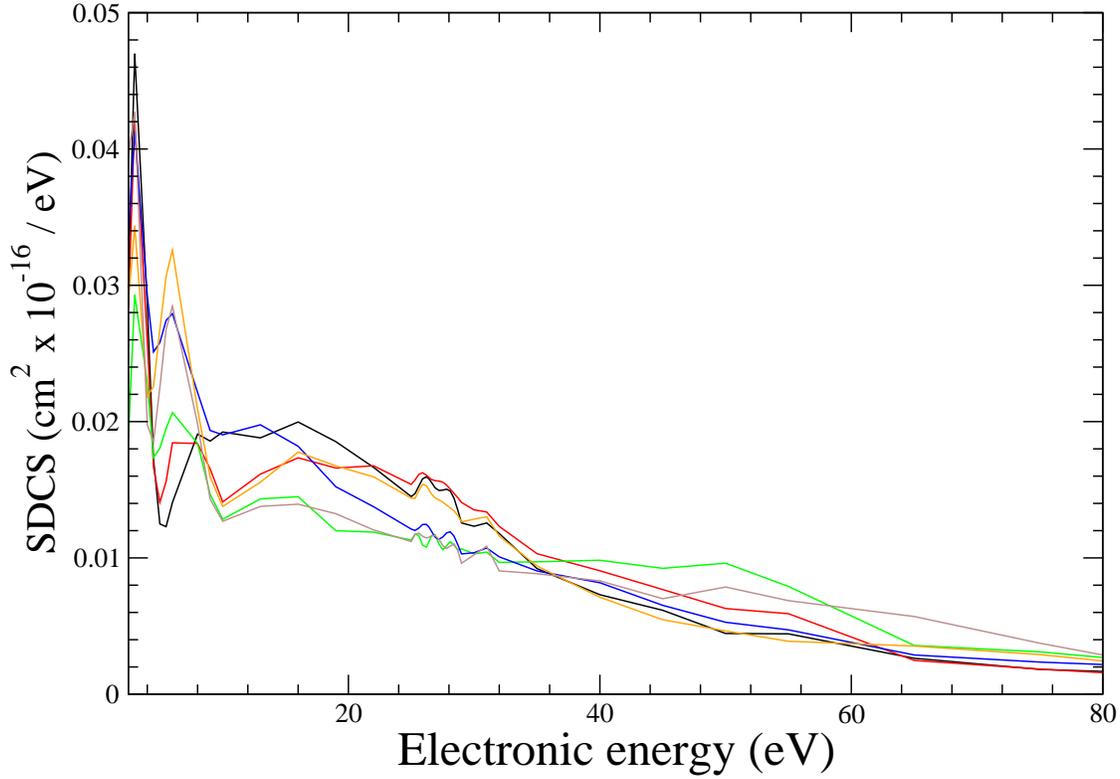}

\centering{}\caption{Single differential cross section for the $p+He$ system as a function
of the emitted electrons energy, calculated at the different distances
$Vt$'s of 100, 125, 150, 175, 190 and 200 a. u. The difference between
the shapes of the curves are due to the phases that multiply the stationary
atomic wave functions. A temporal average of the curves is necessary
to compare the results with data reported in the literature (see Fig.
7).}
\end{figure}

\begin{figure}
\includegraphics[angle=-90,scale=0.6]{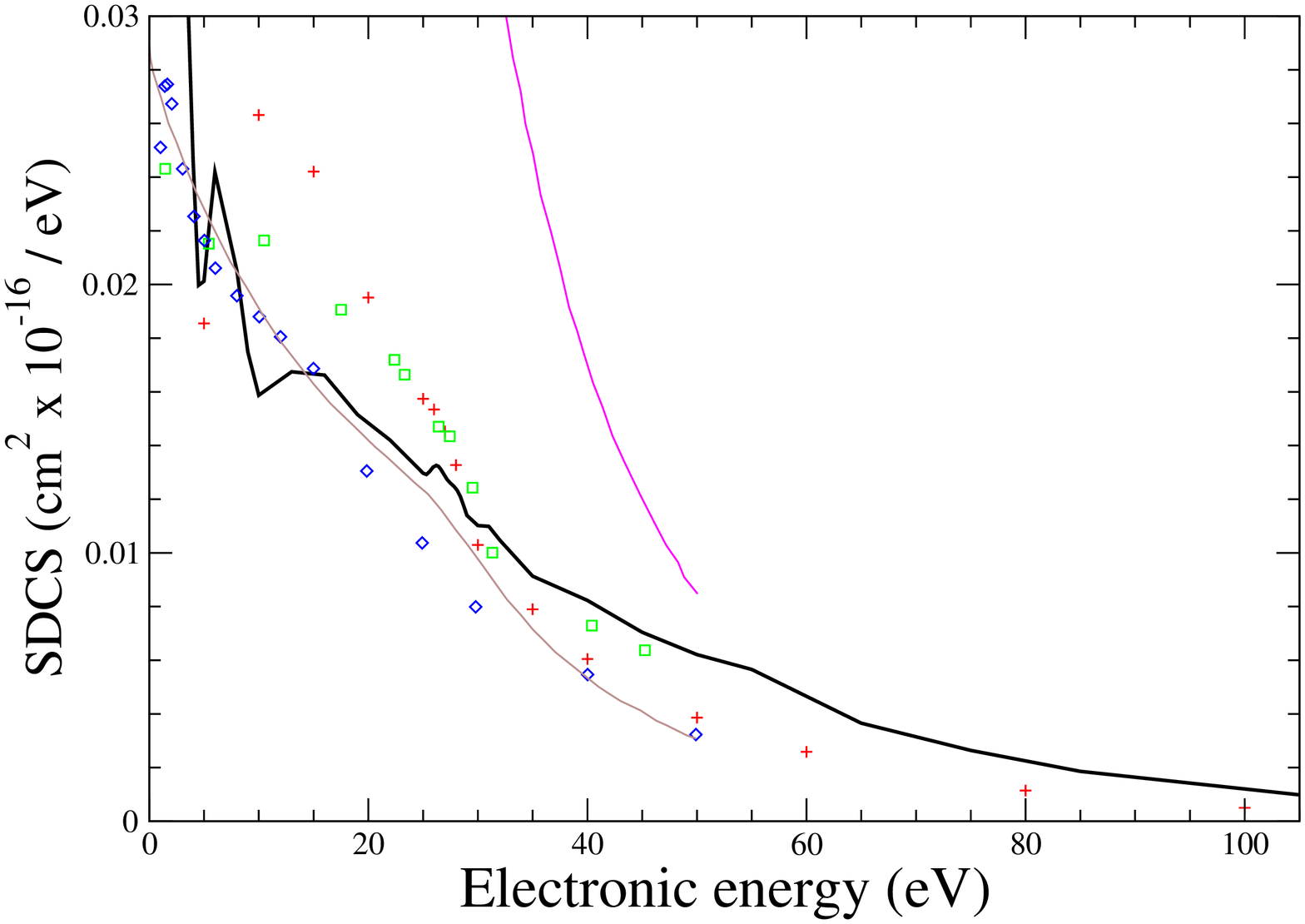}\caption{Ionization distribution in ejected electron energy for the $p+He$
system at 50 keV. For experimental data, the crosses shows data from
Ref. \cite{Gibson:1986}, squares from Ref. \cite{Schulz:1996} and
diamonds from Ref. \cite{Schulz:1996} (where is ascribed to \cite{Cheng:1989}).
For the theoretical calculations, the thin magenta line is Born+PCI
calculations from Ref. \cite{Schulz:1996}, and the brown thin line
is the CWD calculated by Gulyas (Ref. \cite{Schulz:1996}). Considering
20 distributions, the temporal average result found in this work is
given by the thick black line. It shows the general trend of experimental
results above 15 eV.}
\end{figure}

\section{Conclusions }

In this paper the energy distributions of the ionized electrons for
the $p+H$ and $p+He$ systems are calculated at 50 keV. The calculation
is performed in the impact parameter approximation using the method
of Close Coupling in two centers. The results obtained with this method
are compared with previous calculations for these systems, obtained
with similar methods. When a time average on the distributions at
different distances $Vt$ is done, the agreement is reasonable with
some of them and there is an improvement regarding others. It is also
found that the averaged distribution has a betterment in the sense
of becoming softer, and closer to the experimental results as more
curves are included in the average calculation, which we believe is
very important to achieve an agreement with experimental data especially
at low electronic energies. This work is a systematic study that clarifies
ambiguities in other studies regarding magnitudes of final internuclear
distances to be taken into account in the calculation, as well as
suggestions regarding unrealized temporal averages.

As an extension of this work the projectile angle may be incorporated
to obtain double differential cross section, following the approach
studied in Ref. \cite{McGovern:2009}. We do not foresee major complications
inherent to the method for applying this approach to more complex
colliding systems.

\medskip{}

\medskip{}


\section*{Appendix}

\selectlanguage{english}%

\subsection*{Fourier transform of atomic functions }

We can obtain an explicit expression for the Fourier transform of
atomic functions (Eq. \ref{eq:ftr}) following Podolsky and Pauling
\cite{Podolsky:1929}. In terms of spherical coordinates that equation
can be written as: 
\begin{equation}
\begin{array}{rl}
\tilde{\phi}_{lm}\left(\mathbf{p}_{i}\right) & =h^{-3/2}\left(\frac{1}{2\pi}\right)^{1/2}\left[\frac{\left(2l+1\right)\left(l-m\right)!}{2\left(l+m\right)!}\right]^{1/2}{\displaystyle \sum_{k=1}^{k_{max}}}s_{nl}^{k}\left[\frac{\left(2\alpha_{l}\beta_{l}^{k}\right)^{2l+3}}{\left(2l+2\right)!}\right]^{1/2}\\
\\
 & \times\int_{0}^{\infty}dr_{i}\, r_{i}^{l+2}e^{-\alpha_{l}\beta_{l}^{k}r_{i}}\int_{0}^{\pi}e^{-i\,\left(2\pi/h\right)r_{i}p_{i}\cos\theta_{i}\cos\vartheta_{i}}P_{l}^{m}(\cos\theta_{i})\sin\theta_{i}d\theta_{i}\\
\\
 & \times\int_{0}^{2\pi}e^{\pm im\phi_{i}-i\,\left(2\pi/h\right)r_{i}p_{i}\sin\theta_{i}\sin\vartheta_{i}\cos\left(\varphi_{i}-\phi_{i}\right)}d\phi_{i},
\end{array}\label{eq:1A}
\end{equation}
 where $\left(p_{i},\vartheta_{i},\varphi_{i}\right)$ and $\left(r_{i},\theta_{i},\phi_{i}\right)$
denotes the momentum and spatial spherical polar coordinates respectively,
and $h$ is the Planck constant. After the integrations over $\phi_{i}$
and $\theta_{i}$, Eq. \ref{eq:1A} can be simplified as 
\begin{equation}
\begin{array}{cl}
\tilde{\phi}_{lm}\left(\mathbf{p}_{i}\right) & =h^{-3/2}\left(-2\pi\right)\left(-\mathfrak{i}\right)^{l}\left(\frac{h}{p_{i}}\right)^{1/2}\\
\\
 & \times\left(\frac{1}{2\pi}\right)^{1/2}e^{\pm\mathfrak{i}m\varphi_{i}}\left[\frac{\left(2l+1\right)\left(l-m\right)!}{2\left(l+m\right)!}\right]^{1/2}P_{l}^{m}\left(\cos\vartheta_{i}\right)\\
\\
 & \times{\displaystyle \sum_{k=1}^{k_{max}}}s_{nl}^{k}\left[\frac{\left(2\alpha_{l}\beta_{l}^{k}\right)^{2l+3}}{\left(2l+2\right)!}\right]^{1/2}\int_{0}^{\infty}dr_{i}\, r_{i}^{l+3/2}e^{-\alpha_{l}\beta_{l}^{k}r_{i}}J_{l+1/2}\left(\frac{2\pi rp_{i}}{h}\right).
\end{array}\label{eq:2A}
\end{equation}
 For the integral 
\[
I_{kl}\left(p\right)=\int_{0}^{\infty}dr\, r^{l+3/2}e^{-\alpha_{l}\beta_{l}^{k}r}J_{l+1/2}\left(\frac{2\pi rp}{h}\right),
\]
 an analytical expression can be obtained (\cite{Gradshteyn:1965},
p712, eq. 6.623 2.):

\[
I_{kl}\left(p\right)=\frac{2\alpha_{l}\beta_{l}^{k}\left(2\frac{2\pi p}{h}\right)^{l+1/2}\Gamma\left(l+2\right)}{\sqrt{\pi}\left(\left[\alpha_{l}\beta_{l}^{k}\right]^{2}+\left[\frac{2\pi p}{h}\right]^{2}\right)^{l+2}}.
\]
 Substituting this into Eq. $\left[\ref{eq:2A}\right]$, we obtain
\begin{equation}
\begin{array}{cl}
\tilde{\phi}_{lm}\left(\mathbf{p}_{i}\right) & =-h^{-3/2}2\pi\left(-\mathfrak{i}\right)^{l}\left(\frac{h}{p_{i}}\right)^{1/2}Y_{lm}\left(\vartheta_{i},\varphi_{i}\right)\\
\\
 & \times{\displaystyle \sum_{k=1}^{k_{max}}}s_{nl}^{k}\left[\frac{\left(2\alpha_{l}\beta_{l}^{k}\right)^{2l+3}}{\left(2l+2\right)!}\right]^{1/2}\frac{2\alpha_{l}\beta_{l}^{k}\left(2\frac{2\pi p_{i}}{h}\right)^{l+1/2}\Gamma\left(l+2\right)}{\sqrt{\pi}\left(\left[\alpha_{l}\beta_{l}^{k}\right]^{2}+\left[\frac{2\pi p_{i}}{h}\right]^{2}\right)^{l+2}},
\end{array}\label{eq:3A}
\end{equation}
 where $Y_{lm}\left(\vartheta_{i},\varphi_{i}\right)$ are the normalized
spherical harmonics in momentum space. In atomic units, we can rewrite
the previous equation as 
\begin{equation}
\begin{array}{cl}
\tilde{\phi}_{lm}\left(\mathbf{p}_{i}\right) & ={\displaystyle \sum_{k=1}^{k_{max}}}s_{nl}^{k}\frac{-\left(-\mathfrak{i}\right)^{l}2^{2l+3}\left(\alpha_{l}\beta_{l}^{k}\right)^{l+5/2}\left(l+1\right)!}{\left[\pi\left(2l+2\right)!\right]^{1/2}}\frac{\left(p_{i}\right)^{l}}{\left(\left[\alpha_{l}\beta_{l}^{k}\right]^{2}+\left[p_{i}\right]^{2}\right)^{l+2}}Y_{lm}\left(\vartheta_{i},\varphi_{i}\right),\end{array}\label{eq:4A}
\end{equation}
 that when compared to Eq. \ref{eq:cm-1}, defines the functions $F_{l}^{k}$.

\subsubsection*{Acknowledgments }

This work was supported by the Dirección General de Asuntos del Personal
Académico, UNAM, under project IN109511-3.


\begin{thebibliography}{References}
\bibitem{Kirchner:2011} T. Kirchner and H. Knudsen, J. Phys. B: At.
Mol. Phys. \textbf{44}, 122001 (2011).

\bibitem{Arthanayaka:2015} T. P. S. Sharma, B. R. Lamichhane, A.
Hasan, J. Remolina, S. Gurung, L. Sarkadi and M. Schulz, J. Phys.
B: At. Mol. Phys. \textbf{48,} 175204 (2015).

\bibitem{Xiao:2014} F. Xiao-Ying, Z. Rui-Fang, D. Hui-Xiao, S. Shi-Yan
and J. Xiang-Fu, Chin. Phys. B \textbf{23}, 063404 (2014)

\bibitem{Santos:2013} A. C. F. Santos, W. Wolff, M. M. Sant\textquoteright{}Anna,
G. M. Sigaud and R. D. DuBois, J. Phys. B: At. Mol. Opt. Phys. \textbf{46}
075202 (2013).

\bibitem{Rudd:1985} M. E. Rudd, Y. -K. Kim, D. H. Madison and J.
W. Gallagher, Rev. Mod. Phys. \textbf{57,} 965 (1985).

\bibitem{Stolterfoht:1997} N. Stolterfoth, R. D. DuBois and R. D.
Rivarola, \textsl{ Electron emission in heavy ion-atom collisions},
Springer series on atoms and plasmas, Springer, New York (1997).

\bibitem{Ovchinnikov:2004} S. Yu. Ovchinnikov, G. N. Ogurtsov, J.
H. Macek and Yu. S. Godeev, Phys. Rep. \textbf{389,} 189 (2004).

\bibitem{Schulz:1996} M. Schulz, A. Hasan, N. V. Maydanyuk, M. Foster,
B. Tooke and D. H. Madison, Phys. Rev. A\textbf{ 73,} 062704 (2006).

\bibitem{Schultz:2002} D. R. Schultz, C. O. Rienhold, P. S. Krsti\'{c}
and M. R. Strayer, Rev. \textcolor{black}{A}\textbf{\textcolor{black}{{}
65,}}\textcolor{black}{{} 052722 (}2002).

\bibitem{Sidky:1998} E. Y. Sidky and C. D. Lin, J. Phys. B: At. Mol.
Phys.\textbf{ 31,} 2949 (1998).

\bibitem{Chassid:2002} M. Chassid and M. Horbatsch, Phys. Rev. A\textbf{
66,} 012714 (2002).

\bibitem{Fu:2001} J. Fu, M. J. Fitzpatrick, J. F. Reading, and R.
Gayet, J. Phys. B: At. Mol. Phys. \textbf{34,} 15 (2001).

\bibitem{Reading:2005} J. F. Reading, J. Fu, and M. J. Fitzpatrick,
Nucl. Instr. and Meth.\textbf{ B} \textbf{241}, (2005).

\bibitem{Lin:1991} W. Fritsch and C. D. Lin, Phys. Rep. \textbf{20,}
1 (1991).

\bibitem{Reading:1979} J. F. Reading, A. L. Ford, G. L. Swafford
and A. Fitchard, Phys. Rev. A\textbf{ 20,} 130 (1979).

\bibitem{Spranger:2004} T. Spranger and T. Kirchner, J. Phys. B:
At. Mol. Phys. \textbf{37,} 4159 (2004).

\bibitem{Winter:2009} T. G. Winter, Phys. Rev. A\textbf{ 80,} 032701
(2009).

\bibitem{Fitzpatrick:2007} M. J. Fitzpatrick, J. Fu, W. F. Smith,
J. F. Reading, A. Dubois and R. Gayet, Radiation Phys. and Chem. \textbf{76,}
426 (2007).

\bibitem{McLure:1966} G. W. McLure, Phy. Rev. \textbf{148,} 47 (1966).

\bibitem{Kerby:1995} G. W. Kerby, M. W. Gealy, Y. -Y. Hsu, M. E.
Rudd, D. R. Schultz and C. O. Reinhold, Phys. Rev. A\textbf{ 51,}
2256 (1995).

\bibitem{Toshima:1999} N. Toshima, Phys. Rev. A\textbf{ 59}, 1981
(1999).

\bibitem{Amaya:2012} A. Amaya-Tapia and A. Antillón, J. Phys. B,
Conference Series \textbf{388,} 082002 (2012).

\bibitem{Sidky:1999} E. Y. Sidky and C. D. Lin, Phys. Rev. A\textbf{
60,} 377 (1999).

\bibitem{Lee:2007} T-G. Lee, S. Yu. Ovchinnikov, J. Stenberg, V.
Chupryna, D. R. Schultz and J. H. Macek, Phys. Rev. A\textbf{ 76,}
05701 (2007).

\bibitem{Gibson:1986} D. K. Gibson and I. D. Reid, J. Phys. B: At.
Mol. Phys. \textbf{19,} 3265 (1986).

\bibitem{Cheng:1989} W. Q. Cheng, M. E. Rudd and Y. Y. Hsu, Phys.
Rev. A\textbf{ 39,} 2359 (1989).

\bibitem{McGovern:2009} M. McGovern, D. Assafrão, J. R. Mohallem,
Colm T. Whelan and H. R. J. Walters, Phys. Rev. A\textbf{ 79,} 042707
(2009).

\bibitem{Podolsky:1929} W. Q. Cheng, M. E. Rudd and Y. Y. Hsu, Phys.
Rev. A\textbf{ 39}, 2359 (1989). B. Podolsky and L. Pauling, Phys.
Rev. \textbf{34,} 109 (1929).

\bibitem{Gradshteyn:1965} I. S. Gradshteyn and I. M. Ryzhik, \textsl{Table
of integrals, series and products}, Academic Press, New York, (1965).\end{thebibliography}
\end{document}